\documentclass{article}

\usepackage{arxiv}
\usepackage[utf8]{inputenc} 
\usepackage[T1]{fontenc}    
\usepackage{hyperref}       
\usepackage{url}                
\usepackage{booktabs}       
\usepackage{amsfonts}       
\usepackage{nicefrac}       
\usepackage{microtype}      
\usepackage{lipsum}
\usepackage{graphicx}
\usepackage{subcaption}
\usepackage[symbol]{footmisc}

\title{Detecting individual internal displacements following a sudden-onset disaster using time series analysis of call detail records}

\author{
  Tracey Li\thanks{tracey.li@flowminder.org} , Jesper Dejby, Maximilian Albert, Linus Bengtsson, V\'eronique Lefebvre\thanks{veronique.lefebvre@flowminder.org} \\
  Flowminder Foundation\\
  17th May, 2019
  }

\begin{document}
\maketitle
\vspace{1cm}
\begin{abstract}
We present a method for analysing mobile phone call detail records to identify individuals whom we believe to be have been internally displaced as a result of a sudden-onset disaster. We model each anonymous individual’s movement trajectory as a piecewise-constant time series signal, assume that a disaster-induced displacement is exhibited as a level shift from an individual’s ‘normal’ location, and then apply a step detection algorithm to identify level shifts in the signal. In contrast to typical methods that are used to analyse mobility patterns from call detail records, where the aggregate movements of large groups of individuals are analysed, our method offers the advantage that no assumptions regarding the destination or duration of an individual’s displacement are necessary. We have applied the method to the datasets from three disasters - the 2010 earthquake in Haiti, the 2015 Gorkha earthquake in Nepal, and Hurricane Matthew in Haiti in 2016. Our results demonstrate that this method can facilitate improvements in the analysis and modelling of the mobility of internally displaced persons in post-disaster scenarios, using call detail records. Such analyses can be used to complement traditional survey methods to assess the scale and characteristics of disaster-induced displacements in a timely manner.
\end{abstract}


\section{Introduction}

Natural and man-made disasters result in millions of people worldwide being displaced each year. The United Nations High Commission for Refugees (UNHCR) reports that over 68 million people in the world are currently displaced \cite{unhcr2017}. 25 million of these are refugees, meaning that they have crossed international borders. The remainder are internally displaced persons (IDPs), which means that they remain in their home countries under the responsibility of the respective national authority \cite{unhcr2017}.

IDPs are highly vulnerable and the gap between their needs and the resources available to meet them is growing \cite{iom2017}. In many cases, states that have a high number of IDPs may be unable to provide the necessary support to their internally displaced citizens. New approaches to identify and monitor this subset of the population are therefore needed in order to enable their needs to be better met.  

The International Organisation for Migration (IOM) aims to ensure that adequate support and services are provided for displaced persons. Achieving this requires access to accurate data on how many people have been displaced, where they have been displaced from, where they have been displaced to, and the demographic composition of the group. The number of vulnerable persons (children, the elderly, pregnant women, and lactating women) is particularly important. All of this information is difficult, time-consuming, and expensive to obtain; the most common method is for field teams to survey individuals at known displacement sites. However, in many cases the majority of IDPs do not stay at these sites, but instead stay with relatives, with host families, or in rented accommodation. These people are therefore often left uncounted \cite{unhcr2012}. Additionally, it is extremely challenging to obtain accurate information about all the individuals who do stay at displacement sites, because several thousands of people may seek temporary shelter at a site, and enter and leave at different times. There is therefore a need for data collection techniques that can reach beyond the limitations of manual survey methods.

In addition to enumerating the number of displaced persons, effectively responding to crises and anticipating the quantity and location of resources that will be needed is dependent upon an understanding of how people move in times of crisis, with the ultimate goal being the ability to predict how people will move after a shock. It is particularly difficult to collect complete data on the start and duration of a displacement, and data about secondary displacements and onward movements \cite{idmc2017}. This is because the number and complexity of questions that need to be answered in order to reconstruct an accurate picture of someone’s mobility history is high and is reliant on the accuracy of the memory of the respondent. In cases where a displaced person travels between, and temporarily settles at, several different displacement sites over an extended period of time, it is extremely difficult to conduct and piece together longitudinal surveys that accurately capture the person’s trajectory, especially in a post-disaster scenario where the physical and social infrastructure may be severely damaged.

The analysis of mobile phone call detail records (CDR) has been proven to be an effective and valuable way of studying large-scale human mobility after natural disasters \cite{bengtsson2011, lu2012, wilson2016}. These are data that are collected by mobile network operators (MNOs) for billing purposes with one record being generated each time a call or SMS is made or received. The record includes an identifier of the SIM card, the timestamp of the call or SMS, and the location of the cell tower (which is usually the tower closest to the handset) to which the call or SMS was routed. Assuming that each SIM card is used by a single person, this data enables an individual’s spatio-temporal trajectory to be reconstructed, with the location of the cell tower being taken to be the approximate location of the subscriber. As mobile phone penetration continues to increase in all regions of the globe, including in low- and middle-income countries \cite{gsma2017}, CDR analysis is becoming a method through which the movements of large numbers of people can be studied in a meaningful and efficient way; data for a large number of people are available in near real-time and free of interview bias. 

However, CDR data are by no means a perfect dataset. Primarily, the data are most likely not representative of the entire displaced population since only people who own and use a SIM card are included in the data. This typically excludes the youngest and oldest segments of a population, as well as those in the lowest socioeconomic strata. Furthermore, for analysis purposes, it is usually only people who call with a certain frequency that can be included as there are otherwise an insufficient number of data points to perform any meaningful analysis. For these reasons, both the traditional methods of data collection and the analysis of CDR data should be jointly exploited to monitor the movements and distributions of IDPs. Assessing the consistency and differences of the results obtained via the two methods can yield valuable insights into which groups of people are being under-counted or inaccurately represented, and ways in which the methods can be improved.

In this work, we describe a method to analyse CDR data to identify individuals who exhibit a particular pattern of mobility that we believe to be consistent with being internally displaced after a sudden-onset disaster. In contrast to many other CDR analyses of post-disaster mobility that study the collective bulk movements of a population \cite{bengtsson2011, lu2012, wilson2016, bagrow2011}, we instead study the movement patterns of individual subscribers, as in \cite{lu2012, gonzalez2008, song2010}. We apply techniques that are typically used in signal-processing - spatial clustering, temporal filtering, and step detection - to select out individuals that exhibit a movement pattern that we believe to be consistent with disaster-induced displacement. The analysis of this set of subscribers, whom we believe to be IDPs, will be described in subsequent works in which we will also present models of the post-disaster movements of this group.

In the remainder of this paper, we will first describe the context of the three disasters under study and the corresponding datasets. We then proceed to the main body of the work, which is a detailed description of the method used to identify IDPs from a CDR dataset. We briefly present the results of the output of this method and then summarise.

\section{Description of datasets}
\label{sec:datasets}

Access to call detail records was provided by Digicel (Haiti) and NCell (Nepal), both of which are mobile operators that have well-established relationships with Flowminder. Digicel is the dominant telecommunications provider in Haiti, having approximately 4.6 million subscribers out of a total population of over 10 million at the current time, and a market share of over 70\%. In Nepal, NCell has approximately 14.9 million subscribers out of a population of over 29 million, and a subscriber market share of nearly 50\%. The high subscriber counts of these operators mean that the subscriber population is likely to be reasonably representative of the entire national population. 

Table \ref{tab:datasets} summarises the key information for the CDR datasets corresponding to the three disasters being studied.

\subsection{Haiti Earthquake, 2010}
The 2010 earthquake was a catastrophic magnitude 7.0 earthquake that struck on 12th January 2010. The epicenter of the earthquake was approximately 25 kilometres west of Haiti’s capital, Port-au-Prince \cite{usgs2010}. Over the next two weeks, multiple aftershocks were recorded. Major damage was caused in Port-au-Prince, Jacmel, and other cities in the Ouest department. Relief efforts were hampered by the extensive damage to communications, transport, and health infrastructure. It is estimated that 3 million people were affected by the earthquake, including at least 100,000 deaths. Over 1 million people were displaced \cite{kolbe2010}.

The CDR dataset for this time period was successfully used in a study to analyse displacements and population movements in the aftermath of the disaster \cite{lu2012}. Flowminder researchers received data from Digicel comprising the pseudonymised records for the first call of each day for all Digicel subscribers. This dataset covers a period of just over one year from 1st December 2009 until 19th December 2010. The fact that this dataset only includes the first call of the day made or received by each subscriber can limit the extent and accuracy of an analysis. However, a previous study has shown that the location of the first or last call of the day is a good indicator of an individual’s home location during a given time period \cite{bengtsson2011}, which is relevant for this work.

\subsection{Nepal Earthquake, 2015}
The 2015 earthquake, known as the Gorkha earthquake, was a magnitude 7.8 earthquake that struck on 25th April 2015. The epicenter was in the district of Gorkha, approximately 80 kilometres northwest of the capital Kathmandu \cite{usgs2015}. Several aftershocks took place after the main quake throughout the whole country, including one shock of magnitude 6.7. Over 8700 people were killed by the earthquake and an estimated 2.8 million people required humanitarian assistance \cite{reliefweb2015}. The capital city and surrounding districts in the Kathmandu Valley were the most severely affected, but extensive building damage was recorded across the country, with several hundred thousand people being left homeless in several districts \cite{bbc2015}. 

Flowminder received CDR data for this period a few days after the disaster, with a continuous flow of data following. A full voice call dataset is available covering the period between January 2015 and July 2017.

\subsection{Hurricane Matthew, Haiti, 2016}
Hurricane Matthew was a Category 5 storm that caused an estimated 1.9 billion USD of catastrophic damage in Haiti, over 500 deaths, and displaced at least 100,000 families \cite{unicef2016}. The hurricane made landfall over south-west Haiti on the morning of 4th October, 2016. Warnings had been given starting from 1st October. The combination of flooding and high winds caused extensive damage to buildings and infrastructure. The department of Grand Anse suffered the worst effects of the hurricane, followed by the department of Sud. The departments of Nippes, Ouest, Sud-Est, Artibonite, and Nord-Ouest were also affected. 1.4 million people (12.9\% of the population) required humanitarian assistance  \cite{reliefweb2016} and many thousands of people were forced to leave their damaged homes and reside in shelters. The delivery of food, water, and other aid was hampered by road damage, exacerbating the humanitarian crisis.

The CDR dataset for this period spans from March 2016 to May 2017. The data contain all calls made or received by any Digicel subscriber during this period.

\begin{table}
 	\caption{Details of CDR datasets used in this work}
  	\centering
  	\begin{tabular}{llll}
  	\toprule
    			& Haiti Earthquake & Hurricane Matthew & Nepal Earthquake \\
  	\toprule
	Data provider    & Digicel 		      & Digicel     & NCell   \\
	Date of disaster    & 12th Jan 2010     & 4th Oct 2016       & 25th Apr 2015  \\
	Period of data	& Dec 2009 -  Dec 2010 & Mar 2016 - May 2017 & Jan 2015 - Jul 2017 \\
	Daily call volume & Only first call per day  & 2-11 calls per SIM & 2-10 calls per SIM \\
  				& for each SIM is provided & & \\
  	\bottomrule
  	\end{tabular}
\label{tab:datasets}
\end{table}

\section{Method}
\label{sec:method}

\subsection{Overview}
The premise of our method is that a person has been displaced if a level shift is observed in the time series that we call a distance curve. A distance curve is the time series of points representing the distances to someone’s ‘normal’ location. The analysis of such spatio-temporal trajectories is an active field of study, with applications in many fields (see e.g. \cite{zheng2015} for a review). Several of the stages that are commonly required in trajectory analysis - noise filtering, stay point detection, and trajectory segmentation \cite{zheng2015} - are included in our method. 

\begin{figure}
  \includegraphics[width=0.8\linewidth]{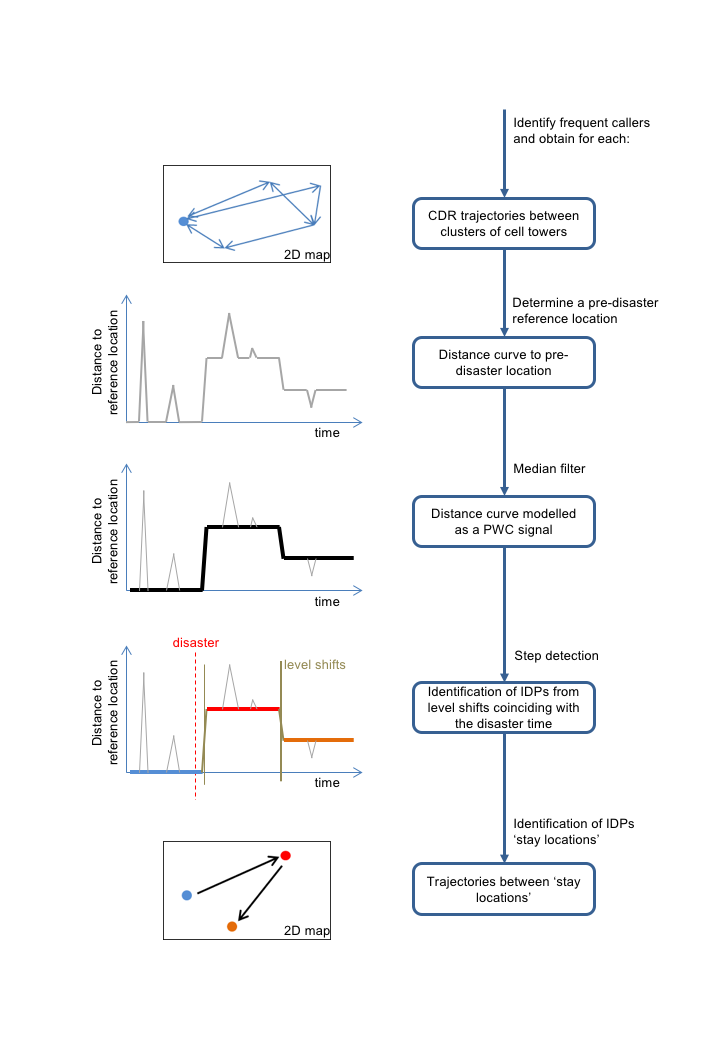}
  \caption{The steps of the method for transforming a CDR time series into a sequence of identifiable `stay locations'. This enables the detection of IDPs by analysing changes in stay locations following a disaster.}
  \label{fig:method}
\end{figure}

Our method begins with transforming the dataset to one that has a satisfactory level of spatial and temporal resolution, by filtering and smoothing. ‘Good’ temporal resolution is ensured by filtering to select a subset of subscribers that meet a specified criteria for calling frequency over the studied time period. The spatial dimension is filtered by applying smoothing via the clustering of cell towers that are ‘close’ together. The steps of the method are then as follows: the two-dimensional sequence of clustered tower locations for each subscriber is reduced into a one-dimensional time series. This is done by first defining a ‘reference location’ for the pre-disaster period and then calculating the scalar distance between that point and the clustered location corresponding to each call. For each day that a call is made, we choose a single distance to use as the data point for that day. We then impute values for the dates where no calls were made to obtain a regularly sampled time series. Filtering is applied to smooth the time series and remove noise in order to create a piecewise-constant (PWC) signal. Finally, a step detection algorithm is applied to this signal to detect the time points at which level shifts have occurred. We then filter out the individuals whom we believe, with high confidence, to have had a stable reference location in the pre-disaster period and identify the subset of those for whom we observe a level shift in the time period immediately following the disaster. These are individuals that we label as IDPs. We identify the locations that these IDPs have stayed at in between level shifts, and transform the CDR time series into a simplified sequences of `stay locations'. The steps of the method are summarised in the diagram in Figure \ref{fig:method}.

Throughout our method, we consider the trade-off between the purity of the IDP subset (i.e. being confident that all individuals in the subset really are IDPs) and the size of the subset. In this initial work, we have chosen to prioritise the purity of the subset and minimise the false positive rate in order to enable us to effectively study the displacement drivers and behavioural patterns that will be presented in subsequent works. This means that several of the filters that we impose are likely to exclude many individuals who are IDPs, and so they will not be included in the analysis subset. We will emphasise, therefore, that the IDP subset that we identify via our method is by no means the complete subset, but is a pure enough sample to enable us to draw confident conclusions in our subsequent analyses.

Another important caveat to highlight is that throughout our work, we make the assumption that one SIM card corresponds to one individual. In practice, it is known that this is not the case as many individuals may own multiple SIM cards, or one SIM card may be shared amongst multiple family members and friends. Developing methods to distinguish and account for these cases is a topic for further study, but has not been included here.

Before moving on to the details of our method, we will briefly discuss here an alternative method for detecting IDPs that we considered, prior to developing the method described in this work. A more generalisable method would be to identify individuals that moved away from home during or immediately after the disaster, but whom we think would otherwise not have done so, based on their previous observed behaviour. This method would be an implementation of an anomaly detection, or change detection, algorithm. Such algorithms are commonly developed and used in the processing of time series for multiple different applications (see, e.g. \cite{dasgupta1996, guralnik1999, liu2013}). They are most effective when the data have high temporal resolution and exhibit some form of regularity that enables a ‘normal’ baseline to be recognised by the algorithm. However, it is very difficult to extract meaningful results from such a method when working with sparse, irregular time series, such as the CDR datasets that we work with in this study. There are extremely few individuals that exhibit any sort of ‘regular’ mobility behaviour and therefore a change detection algorithm picks out nearly all movements to be anomalous. We therefore opted instead to implement a less generalisable and more rule-based method that is better suited to the characteristics of the datasets that we studied.

Details of each of the datasets, and contextual details of each disaster, were provided in Section \ref{sec:datasets}. Here, we briefly discuss the details that are specifically relevant to the methodology and analysis that will be described in the subsequent sections. For each disaster, we would ideally choose to analyse a year of data centred on the disaster date. We have done this where possible. Where it was not possible, because data are missing or because there are problems with data quality, we have chosen instead the closest date that maintains an integer number of weeks. Table \ref{tab:disasters} summarises these details.

\begin{table}
	\caption{Summary details of disaster datasets}
  	\centering
  	\begin{tabular}{llll}
  	\toprule
    			& Haiti Earthquake & Hurricane Matthew & Nepal Earthquake \\
 	\toprule
	Start date (inclusive) & 2009-12-01 & 2016-04-05 & 2015-01-03 \\
	End date (inclusive) & 2010-06-14 & 2017-04-03 & 2015-10-09 \\
	Disaster date &  2010-01-12 & 2016-10-04 & 2015-04-25 \\
	Day of week of disaster & Tuesday & Tuesday & Saturday \\
	Duration of pre-disaster period & 42 days &182 days & 112 days \\
			& (6 weeks) & (26 weeks) & (16 weeks) \\
	Duration of post-disaster period & 182 days & 153 days & 168 days \\
			& (21 weeks) & (26 weeks) & (24 weeks) \\
 	\bottomrule
  	\end{tabular}
  	\label{tab:disasters}
\end{table}

\subsection{Filtering to find frequent callers}
We have investigated and compared two methods for filtering on temporal resolution to identify subscribers who make calls with a consistently high frequency throughout the entire study period. Note that although it would be adequate to only require high calling frequency in the pre-disaster period, we impose the constraint that high frequency is also required for the post-disaster period in order to allow us to perform detailed analyses of the selected subscribers in subsequent studies.

The temporal scale that we have chosen to work at is daily, and therefore this stage of the method is designed to select out subscribers for whom we can obtain adequate temporal resolution on the scale of a day. It is only the number of distinct days on which at least one call is made that is important to us, and not the number of calls made each day, as we will need to assign a daily location to a subscriber on as many days as possible. 

The first method is to study the distribution of call day gaps for each subscriber, and the second is to look at the distribution of the number of call days each week. In both methods, we define a call day to be a day on which a subscriber has made at least one call from an identifiable location. (Locations are not identifiable when, although a cell tower identifier is associated with a call, there is no mapping from that tower to a location and therefore the location of that call cannot be identified). 

\subsubsection{Call day gaps}
A ‘call day gap’ is defined to be the number of days between consecutive call days. If calls are made on consecutive days e.g. Monday and Tuesday, the call day gap is one. If a subscriber makes calls on Monday and then does not call again until Wednesday, the call day gap is two. Days on which we are not confident that we can tell whether people were active are excluded; these are the days in the period immediately following the disaster when many cell towers were non-operational, and days on which data are missing or on which there are a significantly lower number of calls than normal which is indicative of network problems or data quality issues. Excluding a day means that we treat the day before and the day after the excluded period as though they were consecutive days.

By calculating the call day gaps for each subscriber over the observed time period and comparing the distributions between subscribers, we can decide on thresholds for the median and maximum call day gaps such that we can define a ‘good’ degree of temporal resolution that enables us to retain a reasonably large number of subscribers.

\subsubsection{Number of call days per week}
Our second filtering method is to count the number of calls made each week by each subscriber. We define each week to begin on the same day of the week as that of the disaster, which is shown in Table \ref{tab:disasters}. Two criteria are then imposed. The first is to only include subscribers who have made at least one call each week, during every week of the dataset. The second is a cut-off for the median number of call days per week, the value of which is determined by looking at the distribution of call days per week for each subscriber. 

\subsubsection{Comparison of filtering methods}
\label{sec:filtering}
The first method using call day gaps imposes a constraint on the longest time period between data points that is equal to the threshold on the maximum call day gap, and a constraint on the longest time period between 50 percent of the data points that is equal to the threshold on the median call day gap. The second method using call weeks imposes two slightly different constraints, which are restricting the longest time period between data points to be 13 days (which occurs in the case that a call is made on the first day of week 1, and then the last day of week 2), and restricting the longest time period between data points to be 14-\textit{x} days, for roughly 50\% of the time, where \textit{x} is the threshold for the median call days per week.

Although the first method is advantageous because it allows us to control the maximum call day gap to any value, unlike the second one, it is computationally and methodologically more complex to compute call day gaps than it is to count the number of calls in a week. Since we believe that a 13-day gap is acceptable over the duration the studied period, and because we impose stricter constraints around the time of the disaster to ensure high resolution around that period - specifically that there are at least four call days in the week preceding the disaster, at least two call days during the week of the disaster, and at least four calls days in the week following the disaster week, we have chosen to implement the second method using call weeks.

\subsubsection{Filtering results}
\textbf{Haiti Earthquake}: 
In this dataset, there are 2.6 million distinct SIM identifiers of which 26\% have been active during every week of the 6.5 month study period, and 75\% have a median of $\geq$ 6 calls days per week. The number of SIMs that meet both of these criteria and also meet the additional constraints around the disaster period listed in Section \ref{sec:filtering} is 406,546 (16\% of the total).

\textbf{Hurricane Matthew}:
In the 12-month dataset that we have studied (Table \ref{tab:disasters}), there are over 19 million distinct SIM identifiers, many of which only appear for short periods of time. The population of Haiti is around 11 million. This is indicative of a high turnover rate, with many people choosing to buy and use a SIM card for a short period of time. It is also common in low-income countries for people to own multiple SIM cards at a time and to choose to use the one at a particular instance that provides the best economic value. Hence the number of SIM cards that have been used every week may appear to be low - 6\% - but this is in part due to a large fraction of SIM cards being used very infrequently. Of this subset, 66\% have a median of 7 call days per week, and 98\% of those meet the constraints specified in Section \ref{sec:filtering} for the time frame immediately surrounding the disaster. This subset of frequent callers comprises 765,178 SIMs (4\% of all SIMs).

\textbf{Nepal Earthquake}: 
There are over 13.6 million SIM identifiers of which 16\% have registered a call every week and 36\% have a median of $\geq$ 6 calls days per week. There are 1,629,181 SIMs (12\% of SIMs) that meet both these criteria and also meet the additional constraints around the disaster period. These are included in the final subset.

\subsection{Spatial smoothing - clustering tower locations}
\label{sec:clustering}
In general, a caller is connected to the cell tower that is the nearest one to them. Knowing which cell tower a call was connected to and the location of that cell tower, as in a CDR dataset, does not enable the exact point location of the caller to be identified - it is known only that the caller is somewhere ‘near’ the cell tower. If data on the orientations of cell antennae, as well as the cell tower’s location and coverage are available, then probability density maps can be constructed that show the probability of an individual being located at a particular point location e.g. \cite{cbs2018}.

In the absence of such information or if such high spatial precision is not necessary, it is common to localise each individual to the point location of the single cell tower or to construct a Voronoi tessellation of the cell locations and then to assign individuals to a Voronoi cell e.g. \cite{candia2008}. Both of these methods suffer from boundary problems where an individual who is situated midway between two or more towers can be assigned to different towers on different occasions even if they are actually at the same physical location (where ‘location’ is defined to have some finite spatial scale) at all times, because the tower they get routed to is dependent on them making very small movements in either direction. It therefore appears as though the individual is oscillating between two towers, from which it is inferred that they are travelling frequently from one tower to the other, when in reality they may be situated at the same location all the time. 

In this study it is crucial for us to be able to distinguish whether someone is at the same location at which they were previously observed, or if they have moved. Therefore, it is very important for us to address the above boundary problem. We do this by implementing an agglomerative hierarchical clustering method (the ‘fclusterdata’ method from Python’s scipy library with a Ward linkage) to minimise the total within-cluster variance, so that the distance between towers within clusters is minimised and the distance between clusters is maximised. We cluster the point locations (latitude, longitude) of towers using the haversine formula to calculate the great-circle distance between points, with a cutoff threshold that is specified in Table \ref{tab:clustering}. This threshold is chosen to be below the assumed minimal range of a single tower (2 km) and such that the clusters encompass built-up districts such as a village or a city, as assessed from visual comparison with satellite imagery. Clustering towers in this manner significantly reduces the boundary problem but does not fully eliminate it in very dense areas such as city centres, where subscribers can still be observed to oscillate between clusters.

Using this method, many towers in rural locations, where the tower density is very low, remain as isolated single-tower clusters whereas towers in higher-density areas are clustered into groups. The centroid of the convex hull of the tower locations in each cluster is what we refer to as ‘location’ in the remainder of this work.

\begin{table}
 	\caption{Details of spatial clustering of towers}
  	\centering
  	\begin{tabular}{llll}
  	\toprule
    			& Haiti Earthquake & Hurricane Matthew & Nepal Earthquake \\
 	\toprule
 	Clustering threshold & 1 km & 1 km & 1.5 km \\
 	Number of tower sites & 1620 & 710 & 4114 \\
 	Number of clusters & 1102 & 519 & 2535 \\
 	\bottomrule
  	\end{tabular}
  	\label{tab:clustering}
\end{table}

\subsection{Defining a reference location}
A common component of CDR analysis is the determination of a ‘home location’ for each subscriber. There exist several methods to do this that have been described and studied in detail in the literature e.g. \cite{wilson2016, calabrese2012}. In general, ‘home location’ is used to mean the place where a person resides and, in particular, sleeps. Accurately identifying home locations is therefore dependent on having knowledge of local customs and habits to know the standard times during which the population is generally at home. Alternatively, if appropriate survey data is available, then the method can be validated against the survey results.

For the purposes of our study, it is not important for us to identify an individual’s home location as we need only to determine whether or not someone has been displaced from their ‘normal’ location. Therefore, we define a ‘reference location’, which is not necessarily an individual’s home location, but is a location that they commonly visit. We determine the reference location for a specified time period as the location at which a subscriber has been observed on the most days during that time period. In the case that more than one location is the modal location, we choose one at random - this is an arbitrary decision but the choice is relatively unimportant for our purposes. Using this method, we calculate reference locations for all subscribers during the pre-disaster period.

\subsection{Restricting to affected regions only}
To reduce the possibility of mistakenly identifying individuals who have travelled at the same time as the disaster, but not because of the disaster, as IDPs, we restrict our subset to subscribers whose reference location is in one of the administrative 1 regions that were reportedly affected by the disaster. The included regions are shown in Table \ref{tab:regions}.

\begin{table}
 	\caption{List of affected regions for each disaster}
  	\centering
  	\begin{tabular}{llll}
  	\toprule
  	& Haiti Earthquake & Hurricane Matthew & Nepal Earthquake \\
 	\toprule
  	Affected regions & Ouest, Sud-Est & Artibonite, Grande Anse, Nippes, & No filters applied as most  \\
  	 			  &		& Nord Ouest, Ouest, Sud, Sud-Est &  of the country was affected. \\
 	\bottomrule
  	\end{tabular}
  	\label{tab:regions}
\end{table}

\subsection{Calculating scalar distances to reference locations}
We calculate the minimum geodesic distance between each subscriber's reference location and each of the locations at which they have been observed (using the PostGIS ‘ST\_Distance’ function). This reduces the spatial dimensionality of the dataset from two to one.

\subsection{Creating a distance curve for each subscriber}
\label{sec:imputation}
For each day that a call has been made, we choose the minimum distance from the reference location to be the data point for that day. This is because we are interested in whether someone is at their reference location or not for a stable amount of time. If they are seen at their reference location during one day, irrespective of how many other locations they have visited that day, then we assume that the person was at their reference location that day and so their distance is zero.

Summary statistics about calling frequency are shown in Table \ref{tab:calls}. The majority of subscribers did not make a call every day and for those subscribers we therefore have a time series that is sparse and irregular to some degree. We can either choose to work with the irregularly-sampled series, or to impute missing values to create a regularly-sampled series. We choose to do the latter and have considered different options for doing this. These are: back- or forward-filling with the last/next known (or other single) value, back- or forward-filling with an aggregate calculated from a combination of the last/next known values, or guessing the most probable value based on a probabilistic Bayesian analysis of the subscriber’s behaviour. The last option is, in many cases, likely to produce the most accurate results. However, in our case, we are mostly concerned with knowing the person’s behaviour during the disaster period when their behaviour is likely to be unpredictable and not follow the patterns seen during stable periods. Therefore, we choose not to use this method and instead use the second method, implemented as follows:

\begin{enumerate}
	\item Calculate the rolling median over \textit{n} days, using a window to the left.
	\item Fill in the missing rolling median values in the first \textit{n}-1 days of the data using the first non-null value.
	\item Fill in missing dates. Only dates are filled in at this stage - all missing values will be null.
	\item Fill in the missing rolling median values with the nearest previous non-null value.
	\item Fill in the missing time series values with the corresponding rolling median value for that date.
\end{enumerate}

The resulting time series is what we refer to as a `distance curve'.

\begin{table}
 	\caption{Summary statistics about calling frequency. The fraction of subscribers who made a call every day (top row) is low because many cell towers were out of service for some number of days during each of the disasters.}
  	\centering
  	\begin{tabular}{llll}
  	\toprule
   	& Haiti Earthquake & Hurricane Matthew & Nepal Earthquake \\
 	\toprule
  	Fraction of subscribers who 			& 0.08 	& $<$ 0.01 & $<$ 0.01 \\
  	made a call every day 				& 		&		&		 \\
	\midrule
  	Fraction of subscribers who  			& 0.65 	& 0.22 	& 0.30 	\\
  	made a call on at least 95\% of days	&		&		&		\\
	\midrule
  	Fraction of subscribers who 			& 0.93 	& 0.53 	& 0.55 	\\
  	made a call on at least 90\% of days	&		&		&		\\
	\midrule
  	Minimum fraction of call days 		        & 0.64 	& 0.67 	& 0.64 	\\
  	registered by any subscriber			&		&		&		\\
 	\bottomrule
  	\end{tabular}
  	\label{tab:calls}
\end{table}

\subsection{Modelling the distance curve as a piecewise-constant signal}

We use the term ‘noise’ to describe the phenomenon in which a person, in reality, is located stably in the same location (which has a finite physical extension) whereas the time series data make it look as though that person has changed location. This often occurs in situations where there are multiple cell towers close to an individual’s location and the one that they are connected to changes with very short-distance movements. In this case, the individual is seen to ‘oscillate’ between two or more towers. This type of noise is significantly reduced by the spatial clustering of towers described in Section \ref{sec:clustering}.

To reduce this noise further, we apply an iterative median filter, implemented as the `pwc\_medfiltit' function \cite{little2011} with a window of length $n$ days, setting $n=7$. The median filter is a filter that replaces each point in the time series with the median of a window centred at that point. The iterative median filter is an algorithm that repeatedly applies the median filter to a dataset until convergence is reached i.e. until the filtered dataset is identical to the unfiltered one. A median filter is particularly suited to our study because it preserves sharp edges in the signal (level shifts), which are what we want to detect (Section \ref{sec:stepdetection}), unlike Fourier-based signal-processing methods \cite{little2011}. This filter will also have the effect of removing the majority of short trips that are of less than $(n+1)/2$ days’ duration, in which we are not interested. These trips are removed if they are preceded and followed by stable baselines, but may not be filtered out if that is not the case - see Appendix \ref{sec:appendix} for further explanation. A review of alternative noise-reduction techniques can be found in \cite{zheng2015}.

\begin{figure}
  \includegraphics[width=\linewidth]{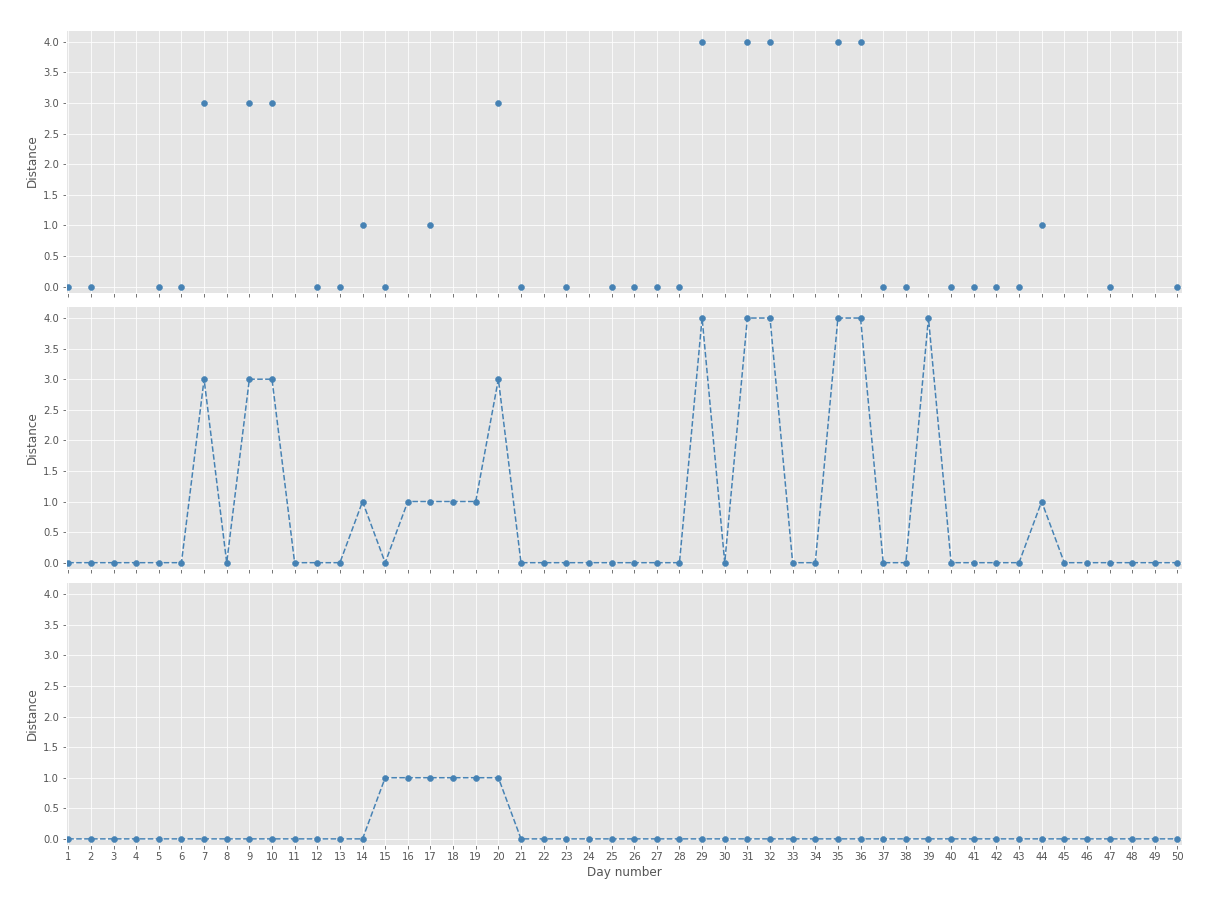}
  \caption{Example of the transformation of a distance curve (top) after missing values have been imputed (middle), and then an iterative median filter applied (bottom).}
  \label{fig:medianfilter}
\end{figure}

Figure \ref{fig:medianfilter} shows an example of how a distance curve (top panel) is transformed first by imputing missing values using the method described in Section \ref{sec:imputation} with a window of length seven (middle panel), and then by the application of the iterative median filter described above with $n=3$ (bottom panel).

The resulting time series is a piecewise-constant (PWC) signal with regularly spaced sampling points. Each piece of the signal is a location at which the subscriber was present for some continuous period of time.

\subsection{Step detection to identify changes in `stay location'}
\label{sec:stepdetection}

We are modelling ‘displacement’ as a sudden shift from one `stay location' to another. Each stay at a location is modelled as a horizontal piece of a piecewise-constant signal, with the vertical ‘steps’, or `level shifts', between pieces representing changes in location. (We use the terms `step' and `level shift' interchangeably from now on). We apply a step detection algorithm to identify the positions of these steps, which are sudden changes in the mean level of the time series. We make the following definitions, where $t$ is the time point of the time series and $x$ is the distance from the reference location at that time point:

$\mbox{diff}(t) = x(t+1) - x(t)$.\\
A step occurs at time $t$ if $\mbox{diff}(t-1) = 0$ and $\mbox{diff}(t) != 0$.\\
The step height is the distance travelled between the start and end locations of the step.\\

Each subscriber’s filtered trajectory is processed to identify the points at which steps occur. The signal for each subscriber is then described completely by a series of constant pieces (stay locations) separated by steps. 

We find that we detect at least one step, at any point in time, for 45\% of subscribers in the ‘frequent caller’ subset for the Haiti earthquake, 55\% of subscribers for Hurricane Matthew, and 67\% for the Nepal earthquake.

\subsection{Identifying potential IDPs from changes in their stay locations}

We impose the following criteria to select individuals who we believe are likely to be IDPs:
\begin{enumerate}
	\item Individuals have a step in the week immediately following the disaster. This seven-day threshold was chosen after examining the figures in Section \ref{sec:idp_steps}, which show that an anomalously high amount of movement (number of steps) occurs in roughly the week following the disaster, and then returns to normal. We also impose a constraint that the stay is of at least three days’ duration, and verify that there were at least two call days away from the reference location during this week\footnote[1]{Due to a processing error, this was implemented as two calls instead of two call days.}. This final constraint is to add confidence that the individual was away for longer than a single day.
	\item Individuals have at least three call days at the reference location in the week immediately preceding the disaster. This is in order to identify departures from the reference location (which can be interpreted as the ‘normal’ or home location) that coincided with the time of the disaster. 
	\item Individuals spent more than half the pre-disaster time at their reference location, as identified from the median-filtered signal. This is so that we can be confident that these individuals were resident at their reference location. The resulting number of IDP candidates is shown in Table \ref{tab:idps}.
\end{enumerate}

\begin{table}
 	\caption{Number of identified IDPs.}
  	\centering
  	\begin{tabular}{llll}
  	\toprule
  	& Haiti Earthquake & Hurricane Matthew & Nepal Earthquake \\
 	\toprule
  	Number of subscribers in 'frequent caller' subset & 765,178 & 406,546 &1,629,181 \\
  	Number of IDPs & 37,839 & 51,070 & 92,806 \\
  	Fraction of frequent caller subset identified as IDPs & 0.05 & 0.13 & 0.06 \\
  	\bottomrule
  	\end{tabular}
  	\label{tab:idps}
\end{table}

\section{Results}
\label{sec:results}
In this section we present a preliminary overview analysis of the subset of individuals that have been identified as IDPs. More detailed analyses will be presented in later works.

\subsection{Comparison of IDP candidates with non-IDP candidates}
\label{sec:idp_steps}
\begin{centering}
\begin{figure}[h!]
  \begin{subfigure}{1.0\linewidth}
    \centering\includegraphics[width=0.75\linewidth]{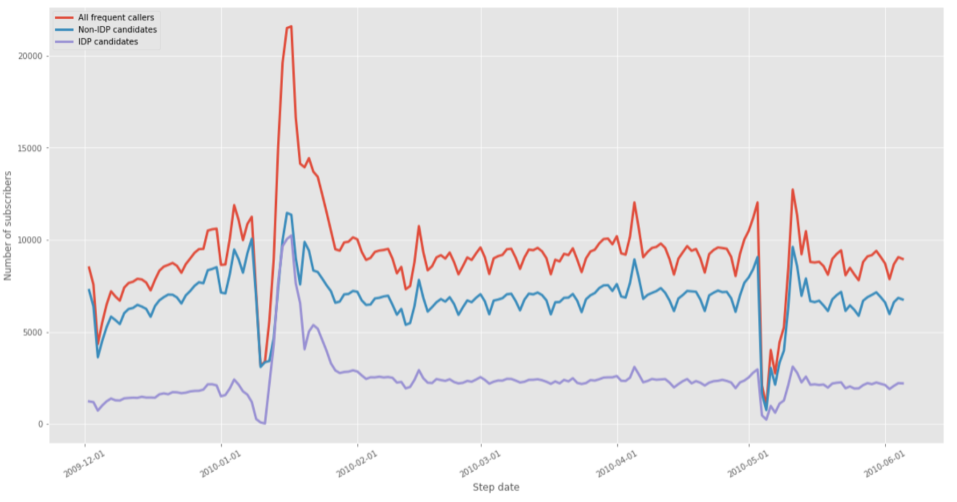}
  \caption{Haiti earthquake} 
  \end{subfigure}
  \begin{subfigure}{1.0\linewidth}
    \centering\includegraphics[width=0.75\linewidth]{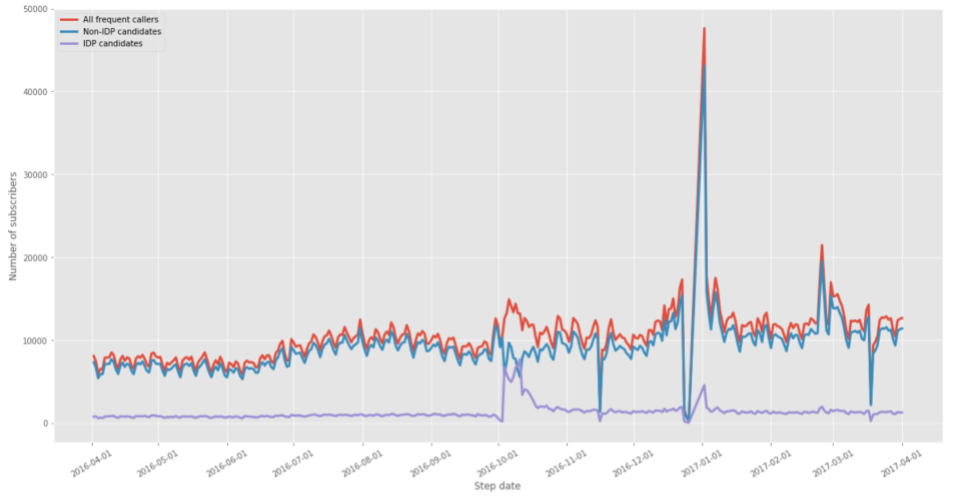}
    \caption{Hurricane Matthew.}
  \end{subfigure}
  \begin{subfigure}{1.0\linewidth}
    \centering\includegraphics[width=0.75\linewidth]{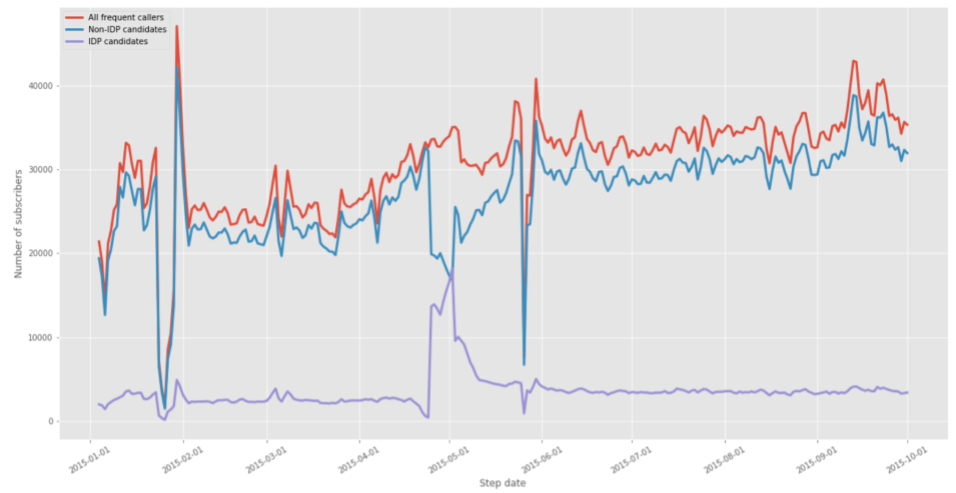}
    \caption{Nepal earthquake.}
  \end{subfigure}
\caption{Number of subscribers making a step each day.}
\label{fig:steps}
\end{figure}
\end{centering}

Figure \ref{fig:steps} shows the number of subscribers from the frequent caller subset who exhibit a level shift each day (red line), divided into the group of subscribers who have been identified as IDPs (purple) and those who have not (blue), for all three disasters. The results we observe are consistent with the differences between IDPs and non-IDPs that we would expect to exist.

\textbf{Haiti earthquake}:
The following trends are visible for all groups: there is a general increase in mobility from December to January, and there is a periodic weekly fluctuation. The large drop at the start of May is due to missing data on a number of days.

For all groups, a sharp drop is visible in the days immediately following the earthquake, which is then followed by a large peak. This is due to several cell towers being out of service in the days following the earthquake which means that no calls could be made or recorded during that time. This pattern is very obvious for the non-IDP group as well as the IDP group, indicating that there are likely to be several subscribers in the non-IDP group who actually are IDPs, but have not been identified by the method.

\textbf{Hurricane Matthew}:
The following trends are visible for all groups: there is a general long-term increase in mobility as the year progresses, there is a periodic weekly fluctuation, and there is a large peak around the Christmas and new year period and a smaller peak at the end of February which coincides with the carnival period.

As expected, there is a sharp and obvious increase in mobility during the disaster period for the IDP group, whereas there is a slight decrease for the non-IDP group. The increase in mobility seen in the post-disaster period, relative to the pre-disaster period, is more abrupt and obvious for the IDP group than the non-IDP group.

\textbf{Nepal earthquake}:
There is a general long-term increase in mobility as the year progresses. The dip and spike at the end of January, and the dip at the end of May, are due to data quality issues.

For the IDP group, there is a sharp and obvious increase in mobility during the disaster period and a decrease for the non-IDP group. This indicates that the effect of the earthquake on many people, who we have not identified as IDPs, was to decrease their mobility. The increase in mobility seen in the post-disaster period, compared to the pre-disaster period, is more abrupt and obvious for the IDP group than the non-IDP group, which could be a sign of prolonged disruption.

\subsection{Fraction of `frequent callers' identified as IDPs in each region}
\begin{centering}
\begin{figure}[h!]
  \begin{subfigure}{1.0\linewidth}
    \centering\includegraphics[width=0.75\linewidth]{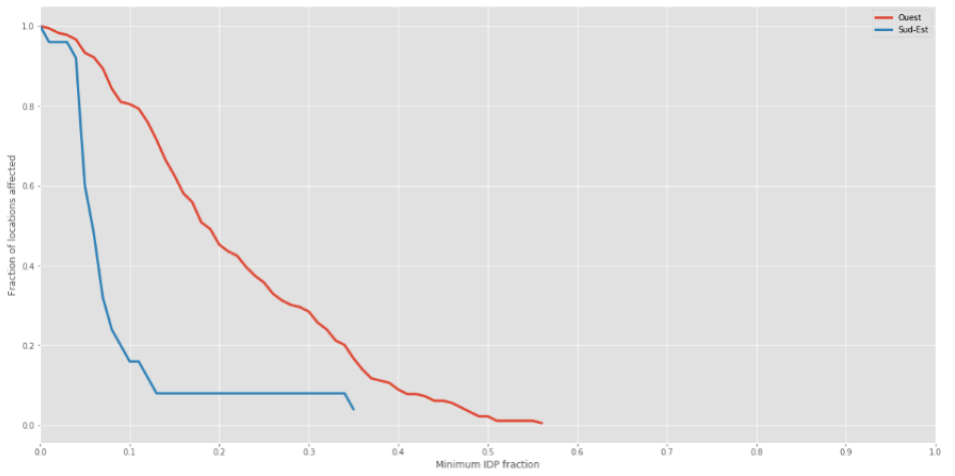}
  \caption{Haiti earthquake} 
  \end{subfigure}
  \begin{subfigure}{1.0\linewidth}
    \centering\includegraphics[width=0.75\linewidth]{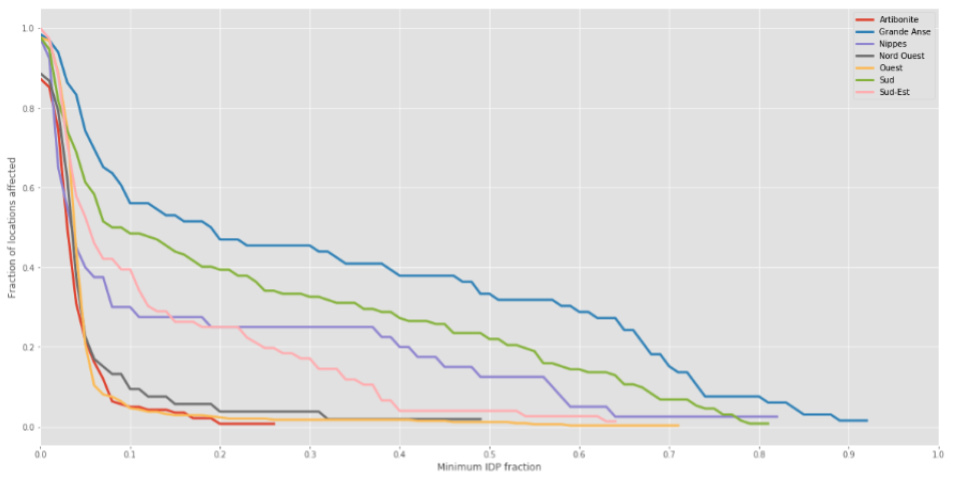}
    \caption{Hurricane Matthew.}
  \end{subfigure}
  \begin{subfigure}{1.0\linewidth}
    \centering\includegraphics[width=0.75\linewidth]{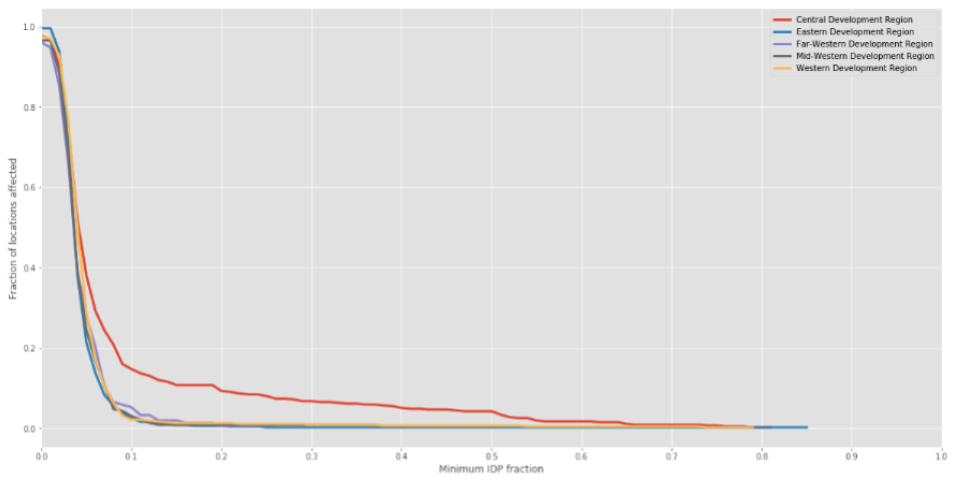}
    \caption{Nepal earthquake.}
  \end{subfigure}
\caption{Fraction of locations affected in each administrative level 1 region as a function of the minimum IDP fraction used to define a location as `affected'.}
\label{fig:affected_clusters}
\end{figure}
\end{centering}

As a further verification that the majority of people that we have identified to be IDPs are likely to really be IDPs that were affected by the disaster, we show in Tables \ref{tab:idps_haiti_earthquake}, \ref{tab:idps_haiti_hurricane}, and \ref{tab:idps_nepal_earthquake} the fraction of frequent callers from each administrative level 1 region that have been identified as IDPs, to check that we observe more IDPs in the regions that were known to be most severely affected. Additionally, we calculate the fraction of frequent callers at each location that have been identified as IDPs, and then calculate the fraction of locations in each region that were ‘affected’, where we define ‘affected’ to mean that greater than a specified minimum fraction of the population was displaced. These graphs are shown in Figure \ref{fig:affected_clusters}. They show that while a small number of IDPs were found in most locations, there are relatively few locations in the regions that reportedly were least affected by the disaster that have a relatively high proportion (e.g. over 30\%) of IDPs. However, in the regions reported to be most severely affected (Grande Anse and Sud for Hurricane Matthew, and Central for the Nepal earthquake), there are many locations with a high proportion of IDPs. In later work, this could be used to calculate a probability (rather than a binary output) that an individual is an IDP, based partly on the behaviour of individuals that share the same home location. 

The results we observe are consistent with the regions that were reported to be the most severely affected having the greatest fraction of IDPs as well as the greatest fraction of affected locations.

\begin{table}[h!]
 	\caption{Haiti earthquake - Fraction of `frequent caller' subset that are identified as IDPs in each administrative level 1 region.}
  	\centering
  	\begin{tabular}{ll}
  	\toprule
  	Administrative 1 region & Fraction \\
 	\toprule
  	Ouest 	& 0.24 \\
	Sud-Est 	& 0.16 \\
  	\bottomrule
  	\end{tabular}
  	\label{tab:idps_haiti_earthquake}
\end{table}

\begin{table}
 	\caption{Hurricane Matthew - Fraction of `frequent caller' subset that are identified as IDPs in each administrative level 1 region.}
  	\centering
  	\begin{tabular}{ll}
  	\toprule
  	Administrative 1 region & Fraction \\
 	\toprule
  	Sud			&	0.20 \\
	Grande Anse	& 	0.15 \\
	Nippes		&	0.11 \\
	Sud-Est		& 	0.07 \\
	Ouest		& 	0.04 \\
	Artibonite		&	0.04 \\
	Nord Ouest	& 	0.04 \\
  	\bottomrule
  	\end{tabular}
  	\label{tab:idps_haiti_hurricane}
\end{table}

\begin{table}
 	\caption{Nepal earthquake - Fraction of `frequent caller' subset that are identified as IDPs in each administrative level 1 region.}
  	\centering
  	\begin{tabular}{ll}
  	\toprule
  	Administrative 1 region & Fraction \\
 	\toprule
  	Central Development	&	0.08 \\
	Western Development	&	0.06	\\
	Far-Western Development & 	0.05 \\
	Eastern Development	&	0.05	\\
	Mid-Western Development &	0.04	\\
  	\bottomrule
  	\end{tabular}
  	\label{tab:idps_nepal_earthquake}
\end{table}

\subsection{Displacement distances}
\begin{centering}
\begin{figure}[h!]
  \begin{subfigure}{1.0\linewidth}
    \centering\includegraphics[width=0.6\linewidth]{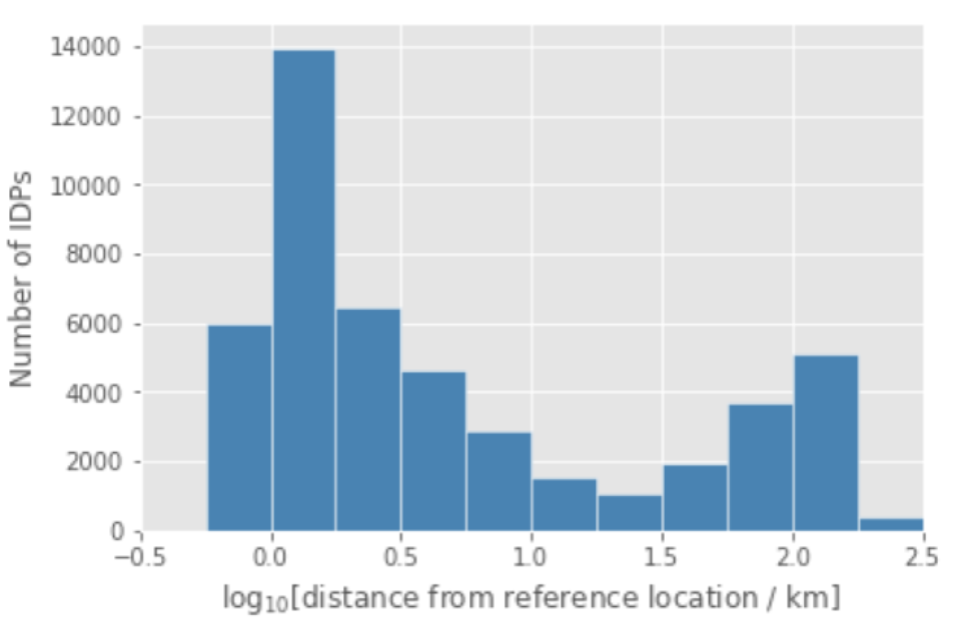}
  \caption{Haiti earthquake} 
  \end{subfigure}
  \begin{subfigure}{1.0\linewidth}
    \centering\includegraphics[width=0.6\linewidth]{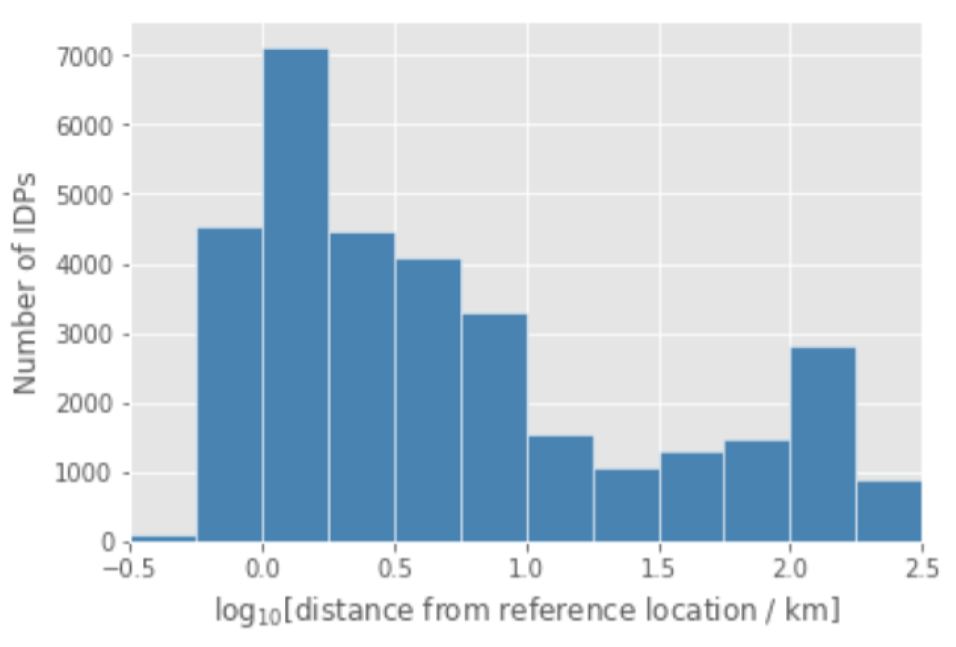}
    \caption{Hurricane Matthew.}
  \end{subfigure}
  \begin{subfigure}{1.0\linewidth}
    \centering\includegraphics[width=0.6\linewidth]{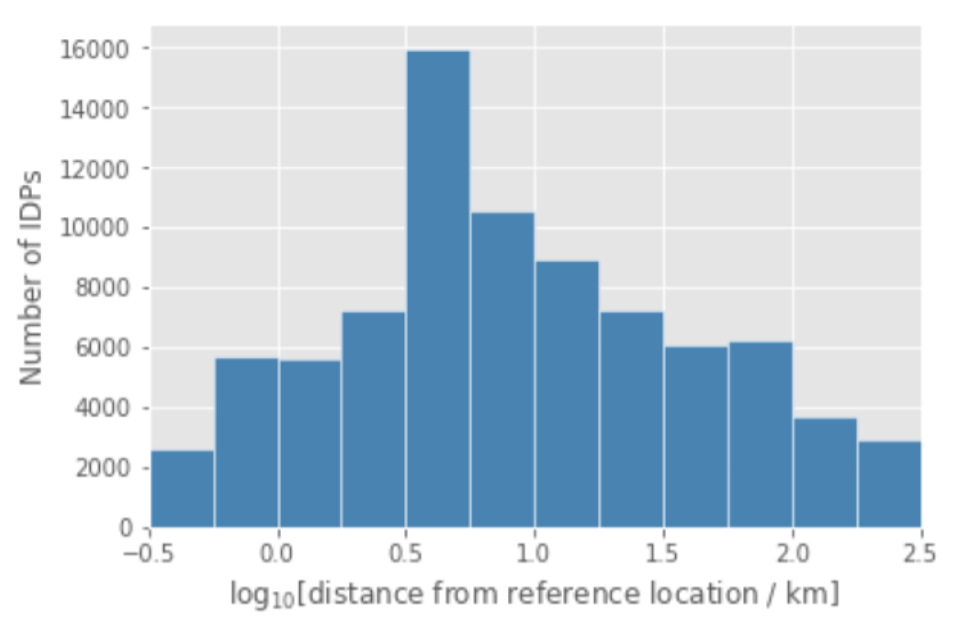}
    \caption{Nepal earthquake.}
  \end{subfigure}
\caption{Distribution of displacement step distances.}
\label{fig:step_lengths}
\end{figure}
\end{centering}

In Figure \ref{fig:step_lengths} we select the first displacement step starting from the pre-disaster reference location that was taken by each IDP in the week immediately after the disaster, and plot the distribution of distances on a logarithmic (base 10) scale. This illustrates that our method is able to identify individuals who were displaced by very short distances (less than a few kilometres), most likely within the same administrative level 3 region. These short-distance displacements are often missed by studies that examine group population movements between regions.

\clearpage

\section{Identification of stay locations between level shifts, for future work}
In subsequent works, we will describe analyses of displacement behaviour that are based on the output of the current work. Knowing where someone has been displaced from and the location that they have been displaced to are crucial inputs to these analyses, and are of the utmost importance to disaster response efforts. 

The output of the present method is the identification of a group of subscribers that are likely to be IDPs. This is accompanied by a sequence of `stay locations’ and dates at which there was a change in the stay location, for each subscriber. This sequence is derived from the decomposition of the individual’s distance curve into a piecewise-constant signal (Section \ref{sec:imputation}). The start and end of each piece, which is where level shifts occur, indicate the times at which we assume there has been a change in location. We use these times to divide up the original CDR data into a series of discrete time periods for each subscriber. For each period, we define a ‘stay location’ to be a location that has been visited on more than one day within that period and is either the unique daily modal location for that period or, if there is no unique daily modal location but one of the modes is the reference location, the reference location is taken to be the stay location. If both of these criteria are not satisfied, we do not assign any stay location to that period.

Following the identification of stay locations for all likely IDPs, we further filter the IDP subset to retain only those for whom we have a high confidence that the pre-disaster location was their most stable location (subscribers who were seen at their pre-disaster reference location at least once in 90 percent of all pre-disaster weeks), and that have at least one identifiable stay location in the post-disaster period that is different to their pre-disaster reference location. This is in order to be able to carry out displacement analyses comparing pre- and post-disaster locations for each identified IDP. After applying these filters, the number of remaining IDPs is shown in Table \ref{tab:idps_final}. These are the IDPs that are used in subsequent analyses which will be described in later works.

\begin{table}[h!]
 	\caption{Final number of identified IDPs used in subsequent analyses.}
  	\centering
  	\begin{tabular}{llll}
  	\toprule
  	& Haiti Earthquake & Hurricane Matthew & Nepal Earthquake \\
  	\toprule
  	Number of IDPs & 32,313 & 47,429 & 78,902 \\
	\bottomrule
	\end{tabular}
  	\label{tab:idps_final}
\end{table}

\section{Summary and discussion}
We have developed a novel method of analysing call detail records (CDRs) for the purposes of identifying the movements of people that we believe have been internally displaced as the result of a natural disaster. The method identifies the subscribers that are likely to be internally displaced persons (IDPs) and the timings of changes in their `stay locations’. This allows for the identification of the locations to which IDPs have been displaced. The resulting IDP subset and their trajectories, described by a sequence of stay locations, can be used in studies to better understand the movements of people in the aftermath of a disaster, and the behaviour of internally displaced persons, via methods that are complementary to traditional survey methods. The main distinguishing feature of our work, relative to other disaster response studies using CDR data, is the analysis of individual movement trajectories rather than group movements between two locations. This will allow for a much deeper understanding of post-disaster behaviour that can provide valuable input to inform disaster-preparedness and disaster-response efforts. For example, it allows us to study people that have been displaced over very short distances, of less than a few kilometres, who are often missed in studies of group population movements. Our method also does not require us to make any assumptions on where an individual has been displaced from or to. It is important to stress that although we analyse individual-level movements, any results and insights that are presented contain only aggregated data from which it is impossible to identify an individual. Additionally, we have ensured that the processing of all personal data has been performed in accordance with high standards of data governance and information security best practices.

Our method can facilitate advances in the modelling and understanding of human mobility in post-disaster scenarios, using mobile operator data or similar spatio-temporal datasets. Due to the characteristics of ‘big data’ - these datasets are automatically generated and contain records for millions of people - these data can yield insights on a scale, in a timeframe, and with a precision that is unfeasible using data that are collected manually, such as displacement-site surveys that are a common source of information during a humanitarian crisis.

We recognise the limitations of the method we have developed. Our method filters out individuals based on examining specific features of their mobility behaviour. We have been deliberately conservative throughout the entire method in order to minimise the number of false positives - individuals whose movements match the patterns we expect from IDPs, but who are not IDPs. This means that we have a very high rate of false negatives - people who are IDPs but who we do not identify as such. We have chosen to do this because it is important for the analyses and models that will follow on from this work and that will be described in later papers. However, we would like to perform follow-up studies to estimate the true/false positive/negative rates obtained by our method. From this, we would then be able to estimate the number of IDPs in the entire, unfiltered dataset.

It is important also to mention the inherent limitations of a CDR dataset. Primarily, the dataset only contains records for people that use a mobile phone and the data are therefore unlikely to be representative of the entire population. In particular, the youngest and oldest members of the population, and individuals in the lowest socio-economic strata will likely not be represented. Whilst this is not something that needs to be taken into consideration at this stage of our work, we will discuss the issue more fully in later studies.

We hope that this work will stimulate discussions and the further development of methods to improve crisis-response efforts. The authors appreciate feedback from interested parties.

\section{Acknowledgements}
This work was performed as part of the project ‘Contributing to a better understanding of human mobility in crisis and enhancing linkages with citizen-driven assistance’, a collaborative project between the International Organisation for Migration (IOM), the Internal Displacement Monitoring Centre (IDMC), the Humanitarian Data Exchange (HDX, part of the United Nations Office for the Coordination of Humanitarian Affairs (OCHA)), and Flowminder, funded by the European Civil Protection and Humanitarian Aid Operations (ECHO). Access to the de-identified mobile operator data used in this work was generously provided by Digicel (Haiti) and NCell (Nepal). Processing of that data was performed whilst adhering to high levels of data governance and information security best practices.

\bibliographystyle{unsrt}  
\bibliography{references}  

\appendix
\section{Iterative median filter}
\label{sec:appendix}
The ‘pwc\_medfiltit’ python function \cite{little2011} is a method that repeatedly applies the ‘medfilt’ function from Python’s scipy library to a dataset until convergence i.e. until the filtered data is the same as the unfiltered data. A single application of the ‘medfilt’ function with a window of length $n$ replaces all values of a dataset with the median value of a window of length $n$ centred on each value. 

In the context of our work, where we are interested in detecting shifts from a baseline of zero, any non-zero sequences of length $< n$ will be filtered out if they are surrounded by a ‘clean’ baseline. Sequences of length $\geq n$ will remain as non-zero sequences although the values in that sequence may change. However, from our perspective, the crucial point is that the presence of a non-zero sequence has been detected. Below are some examples, using $n=3$.

\textbf{Example 1}: sequence of length 2 surrounded by ‘clean’ baseline \\
Unfiltered data: 	$[0, 0, 3, 1, 0, 0, 0, 0]$ \\
Filtered data:	  	$[-, 0, 0, 0, 0, 0, 0, 0]$

\textbf{Example 2}: sequence of length 2 surrounded by ‘noisy’ baseline \\
Unfiltered data:	$[0, 0, 3, 1, 0, 2, 0, 0]$ \\
Filtered data:		$[-, 0, 1, 1, 1, 0, 0, 0]$

\textbf{Example 3}: sequence of length 3 surrounded by ‘clean’ baseline \\
Unfiltered data: 	$[0, 0, 3, 1, 2, 0, 0, 0]$ \\
Filtered data:		$[-, 0, 1, 2, 1, 0, 0, 0]$

\end{document}